\documentclass[conference]{IEEEtran}
\IEEEoverridecommandlockouts
\usepackage{cite}
\usepackage{amsmath,amssymb,amsfonts}
\usepackage{algorithmic}
\usepackage{graphicx}
\usepackage{textcomp}
\usepackage{subcaption}
\usepackage{xcolor}
\usepackage{xspace}
\usepackage{booktabs}
\usepackage{url}
\def\BibTeX{{\rm B\kern-.05em{\sc i\kern-.025em b}\kern-.08em
    T\kern-.1667em\lower.7ex\hbox{E}\kern-.125emX}}

\newcommand{\mypar}[1]{\vspace{2mm}\noindent\textbf{#1}}
\def\etal{\textit{et al.}\xspace}

\usepackage{colortbl}
\usepackage{todonotes}
\usepackage{comment}

\begin{document}

\title{%
    Detecting Audio Deepfakes on the Edge: Lightweight SSL-Based Detection in a Browser Plugin
    \thanks{%
        This work was supported by the EU Horizon project AI4TRUST (No. 101070190) and by a grant of the Ministry of Research, Innovation and Digitization, CNCS/CCCDI - UEFISCDI, project number PN-IV-P7-7.1-PTE-2024-0600, within PNCDI IV.
    }
}

\author{
\IEEEauthorblockN{%
    Octavian Pascu\IEEEauthorrefmark{1},
    Dan Oneata\IEEEauthorrefmark{1},
    Horia Cucu\IEEEauthorrefmark{1},
    Nicolas M. Müller\IEEEauthorrefmark{2}%
}
\IEEEauthorblockA{\IEEEauthorrefmark{1}%
National University of Science and Technology \textsc{Politehnica} Bucharest, Romania\\
\texttt{\{octavian.pascu,dan\_theodor.oneata,horia.cucu\}@upb.ro}}
\IEEEauthorblockA{\IEEEauthorrefmark{2}%
Fraunhofer AISEC, Germany \,\, \texttt{nicolas.mueller@aisec.fraunhofer.de}}
}

\IEEEoverridecommandlockouts
\IEEEpubid{\makebox[\columnwidth]{979-8-3315-7486-4/25/\$31.00~\copyright2025 IEEE \hfill}
\hspace{\columnsep}\makebox[\columnwidth]{ }}

\maketitle
\IEEEpubidadjcol
\begin{abstract}

Audio deepfakes are a growing challenge for the general public, as well as for journalists and fact-checkers.
The latter need reliable tools to verify the authenticity of their sources, while at the same time keeping their information private.
Commercial deepfake detection solutions rely on cloud-based processing, which raises privacy concerns.
To solve this problem, we propose an on-device audio deepfake detection model.
We show that a truncated self-supervised backbone with a simple logistic classifier is both very fast and often more accurate than existing solutions.
Our solution outperforms the baseline AASIST by 10\% and improves inference speed by 40\%.
We integrate this model into a browser plug-in,
which allows journalists and fact-checkers to detect deepfakes easily and securely. Code for the plugin is available at \url{https://github.com/OctavianPascu97/Audio-Deepfakes-Browser-Plugin}.

\end{abstract}

\section{Introduction}

Text-to-speech systems have achieved remarkable progress in recent years and are now capable of synthesizing human-like speech with high fidelity.
This progress has enabled transformative applications in entertainment, accessibility, and creative fields.
At the same time, generative AI poses a threat to the society at large:
each of us can be deceived by highly convincing synthetic samples (also known as deepfakes) that are designed to spread fake news or misinformation.

To counter this growing threat, researchers have been actively developing detection systems capable of identifying synthetic speech. However, for such systems to be widely adopted—particularly by journalists and fact-checkers—they must meet several key criteria: (i) high accuracy, (ii) strong privacy protections, and (iii) ease of use, with seamless integration into everyday applications and minimal computational requirements.

Existing audio deepfake detection solutions face several limitations. Many open-source tools \cite{jung2022aasist, 
jung2022pushing,
pascu2024towards, 
tak2021end, 
tak2020end, 
tak2022automatic}
are trained on controlled datasets but fail to generalize to real-world deepfakes. Additionally, these tools often require technical expertise and substantial computing power, limiting their accessibility.
Moreover, cloud-based solutions 
\cite{Deepfake78:online, SensityA49:online}
require users to upload audio samples for analysis, raising concerns about data security and privacy—an especially critical issue for journalists and fact-checkers who must protect sensitive sources.

\begin{figure}[!t]
    \centering
    \begin{subfigure}{0.39\linewidth}
        \centering
        \includegraphics[width=0.9\linewidth]{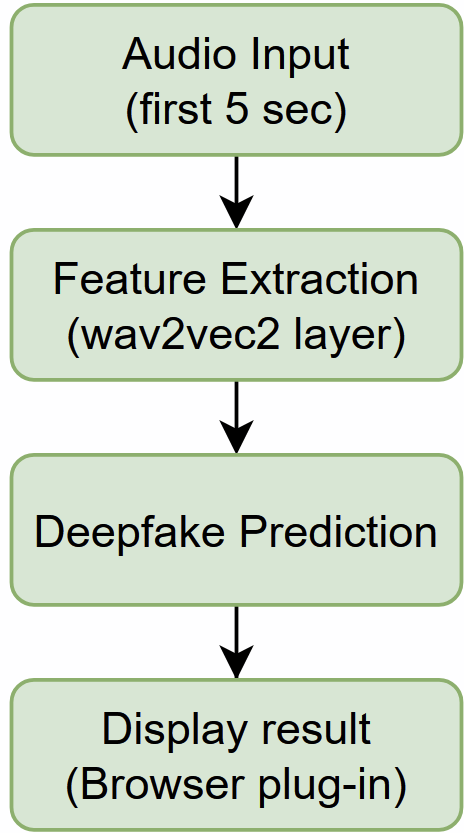}
        \caption{Detection pipeline.}
        \label{fig:pipeline}
    \end{subfigure}
    \hfill
    \begin{subfigure}{0.59\linewidth}
        \centering
        \includegraphics[width=0.9\linewidth]{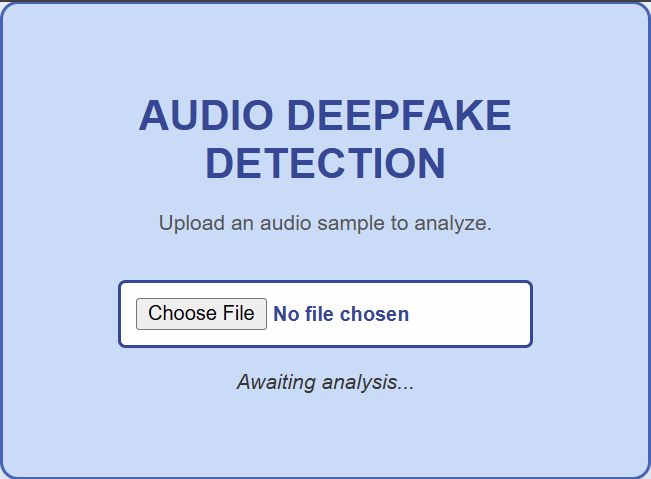}
        \caption{Chrome extension UI.}
        \label{fig:ui}
    \end{subfigure}
    \caption{Audio Deepfake Detection Chrome Extension: (a) system pipeline; (b) user interface.}
    \label{fig:UI}
\end{figure}
In this paper, we introduce a browser extension capable of performing real-time audio deepfake detection directly on consumer devices.
Since the detection runs locally, our solution ensures complete user privacy.
To achieve both accuracy and efficiency, we leverage state-of-the-art self-supervised learning (SSL) representations.
Instead of employing the full network, we truncate it to the initial layers and apply a lightweight linear classifier.
A systematic evaluation on six deepfake detection datasets shows that our truncated model outperforms the full network, achieving higher accuracy while reducing computational costs.
Our approach also compares favorably to other more complex methods in terms of both accuracy and efficiency.

To summarize, \textbf{our main contributions} are as follows. First, we propose a lightweight yet effective approach for audio deepfake detection, leveraging pre-trained SSL features and a linear classifier. Second, we evaluate our method on six out-of-domain datasets, achieving an average relative error reduction of 25\% while improving efficiency in terms of speed and memory usage (inference speed reduced by 40\%). Third, we integrate our model into a user-friendly Chrome extension, making deepfake detection accessible to the general public.

\section{Related Work}  

Current efforts, such as the ASVspoof Challenge, focus on improving \textbf{detection performance} under controlled settings. However, in real-world scenarios, two additional factors are equally important: \textbf{generalization}, the ability of a model to detect previously unseen deepfake generation techniques; and \textbf{computational efficiency}, the balance between model size, inference speed, and energy consumption, which directly impacts deployability.

\subsection{General considerations}

\subsubsection{The Challenge of Generalization}  

Ensuring robust generalization remains a major challenge in deepfake audio detection, as models often fail when encountering previously unseen attack methods. Müller et al. \cite{muller2024harder} systematically analyzed why deepfake detection models struggle to generalize across different datasets and synthesis techniques. Their study decomposed generalization failures into hardness and difference components, showing that existing detection models overfit to specific attack artifacts, leading to a sharp performance drop on unseen forgeries. 
In our study, we focus on generalization as a performance measure by evaluating detectors across six datasets comprising samples produced by a diverse range of text-to-speech and vocoder architectures.

\subsubsection{Computational Efficiency and Practical Deployment}  

Beyond generalization, model size and inference efficiency are critical for deployability in real-world applications. 

Larger models generally achieve higher detection accuracy, as shown by Pascu et al. \cite{pascu2024towards}. However, models with billions of parameters are impractical for deployment in most real-world applications. For instance, an SSL model with two billion parameters may be effective in research settings, but its computational demands make it unusable in mobile or real-time environments.  

\subsubsection{Privacy, Ease of Use}  

Beyond security concerns, deepfake detection solutions should also prioritize privacy and ease of use:  

\noindent\textbf{Privacy.} Many existing detection systems rely on cloud-based processing, requiring users to upload audio for analysis \cite{Deepfake78:online,Deepfake50:online,SensityA49:online}.
This poses risks related to data security and user privacy. 
In contrast, we propose a lightweight on-device solution that can perform detection without transmitting sensitive speech data.

\noindent\textbf{Ease of Use.} A practical detection system should be easy to integrate into everyday applications, requiring minimal computational resources while remaining effective. To this end, we propose a lightweight detection system that runs locally as a browser extension. Aside from installing the browser itself, no additional technical setup is required to use the application.

\subsection{Architectures and models}
In this subsection we describe models and architectures related to audio deepfake detection (including models that we compare our approach to).

\subsubsection{RawNet and Graph-Based Approaches}  
Deepfake audio detection has evolved significantly with the development of raw waveform-based deep learning models. RawNet2 \cite{tak2020end} is an end-to-end convolutional neural network (CNN) that directly processes raw audio waveforms, eliminating the need for handcrafted spectral features. It has been widely used in deepfake detection, achieving high performance in speaker verification and synthetic speech detection tasks. However, recent studies have highlighted its limitations in handling unseen spoofing attacks and its susceptibility to adversarial perturbations \cite{tak2021end}.  

To address these limitations, RawGAT \cite{tak2021end} was introduced as an evolution of RawNet2, incorporating graph attention networks (GATs) to enhance the model’s ability to capture spectro-temporal dependencies in audio signals. By leveraging graph-based structures, RawGAT improves the detection of subtle deepfake artifacts, making it more robust against variations in synthetic speech generation techniques.

\subsubsection{Spectro-Temporal Models for Deepfake Detection}  

AASIST (Audio Anti-Spoofing using Integrated Spectro-Temporal Graph Attention Networks) \cite{jung2022aasist} combines RawNet2 with spectro-temporal graph attention layers, integrating both temporal and spectral features of speech signals. Unlike RawNet-based models, which primarily rely on raw audio feature extraction, AASIST enhances deepfake detection by leveraging multi-scale spectro-temporal dependencies, making it particularly effective against advanced TTS and voice conversion (VC) attacks. 

\subsubsection{Self-Supervised Learning for Deepfake Detection}  

Self-supervised learning (SSL) models, particularly Wav2Vec2 \cite{pascu2024towards, tak2022automatic}, have demonstrated significant advancements in deepfake detection by leveraging large-scale unlabeled speech data to learn robust representations. 
Unlike conventional models that rely on predefined feature extraction, SSL models hierarchically capture fine-grained distortions introduced by deepfake generation techniques, making them well-suited for spoofing detection.  

Pascu et al. \cite{pascu2024towards} explored the application of self-supervised representations for cross-dataset generalization in deepfake detection. Their study revealed that using frozen SSL representations with a simple classifier significantly improved performance, reducing the Equal Error Rate (EER) from 30.9\% to 8.8\% across eight deepfake datasets. Furthermore, they emphasized the importance of model calibration, ensuring that SSL-based deepfake detectors produce reliable confidence scores, enhancing trustworthiness in real-world applications.  

Tak et al. \cite{tak2022automatic} investigated the use of Wav2Vec2 for automatic speaker verification spoofing detection, fine-tuning it to distinguish between bona fide and synthetic speech. Their results indicated that SSL-based models achieved state-of-the-art performance, even when trained exclusively on bona fide speech samples. Additionally, they introduced RawBoost, a data augmentation framework that applies adaptive filtering, dynamic range clipping, and reverberation simulations to enhance robustness against real-world distortions and attack conditions.  

Layer-wise analysis of large SSL models has shown that different layers encode different acoustic properties \cite{pasad2021layer}.
Guo et al. \cite{guo2024audio} investigated the impact of feature extraction layer selection in SSL-based deepfake detection. Their findings indicate that extracting features from mid-to-lower layers significantly improves performance, as these layers retain fine-grained distortions introduced by deepfake synthesis techniques. 

\section{Methodology}

Our goal is an audio deepfake detection model that works well, but is also lightweight,
such that it can be deployed on consumer devices.
Self-supervised representations (such as \textit{Wav2Vec2}) remain the de facto frontend for deepfake detection \cite{tak2022automatic, pascu2024towards} 
but these are generally heavyweight;
for example, Pascu \etal argue for using the two-billion parameter \textit{Wav2Vec2} model \cite{pascu2024towards}.
We propose to make the \textit{Wav2Vec2} frontend more efficient by truncating it and using the first few layers only, instead of all the layers.
To ensure efficiency in the backend, we use a linear classification layer on top.
This was shown to be give strong generalisation capabilities \cite{pascu2024towards}.

\mypar{Wav2Vec2 representations.}
Self-supervised audio representations encode high-level acoustic and phonetic patterns and have proven useful in many tasks such as speech recognition or speaker identification.
They achieve this by training on large quantities of raw audio and predicting masked parts of the input audio file.
Here we use the \textit{Wav2Vec2} family \cite{baevski2020wav2vec} and in particular the XLS-R-300M variant \cite{babu2021xls},
because it was shown to strike a good tradeoff between accuracy and efficiency \cite{pascu2024towards}.
This model was pre-trained on multilingual speech (CommonVoice, Babel, Multilingual LibriSpeech, VoxPopuli and VoxLingual107) and totals 310 million (M) parameters.
Its architecture consists of a convolutional frontend (0.5M parameters), followed by 24 Transformer layers (each with around 13M parameters).
We extract features from the frozen model after each individual layer:
from the first up to the the last (the 24th) transformer layer.
We use the pretrained model, with no fine-tuning.

\mypar{Linear classifier.}
On top of the extracted features, we train a linear classifier to predict whether the audio file is real (bona fide) or fake (spoofed).
The classifier is learnt by optimizing the binary cross entropy loss plus an L2 regularization term.
The model is implemented using logistic regression from \texttt{scikit-learn} using 5,000 maximum iterations, $C=10^6$, and the rest default parameters.
Given that the \textit{Wav2Vec2 XLS-R-300m} model produces 768-dimensional features,
there are 769 learnable parameters: 768 weights corresponding to the features and one bias term.
To assess the performance of the classifier, we use the equal error rate (EER), which is a threshold-free metric.

\mypar{Datasets.}
We train the linear classifier on the ASVspoof19 \cite{wang2020asvspoof} dataset and, in order to assess the generalization capabilities,
we evaluate it on a combination of six other datasets,
which are described in Table~\ref{tab:dataset_summary}.

\begin{table}[ht]
    \centering
    \small
    \renewcommand{\arraystretch}{1.2}
    \setlength{\tabcolsep}{3pt}
    \begin{tabular}{@{}lccr@{}}
        \toprule
        \textbf{Dataset} & \textbf{Languages} & \textbf{Systems} & \textbf{Utterances} \\
        \midrule
        ASVspoof 2019 \cite{wang2020asvspoof}  & English      & 19   & 121k  \\ 
        ASVspoof 2021 DF \cite{yamagishi2021asvspoof} & English      & 100+  & 593k  \\ 
        Fake Or Real (FoR) \cite{reimao2019dataset}& English      & 7 & 195k       \\ 
        MLAAD \cite{muller2024mlaad}         & 38 Lang.     & 82  & 154k \\ 
        In the Wild (ITW)   \cite{muller2022does}        & English      & N/A  & 31k       \\ 
        TIMIT  \cite{salvi2023timit}    & English      & 12   & 20k    \\ 
        WaveFake  \cite{frank2021wavefake}     & English, Japanese      & 9   & 136k      \\ 
        \bottomrule
    \end{tabular}
    \caption{Overview of Audio Deepfake Detection Datasets}
    \label{tab:dataset_summary}
\end{table}

\mypar{Augmentation.}
All our experiments were done using RawBoost, a data augmentation technique designed to enhance deepfake audio detection by simulating real-world distortions. It applies additive noise, reverberation, equalization, and compression to raw audio waveforms, making deepfake detection models more robust to manipulated speech. 
The configuration used for RawBoost is the one proposed by Tak \etal \cite{tak2022automatic} for the LA partition.

\section{Experimental results}

\subsection{Layer-wise analysis of Wav2Vec2 for Deepfake Detection}
\label{subsec:layerwise-analysis}

\begin{table}[!ht]
\setlength{\tabcolsep}{1pt}
\begin{tabular}{r|rrrrrrr||r}
\toprule
\multicolumn{1}{c}{\textbf{Layer}} & \multicolumn{1}{c}{\textbf{ASV19}} & \multicolumn{1}{c}{\textbf{ASV21}} & \multicolumn{1}{c}{\textbf{FoR}} & \multicolumn{1}{c}{\textbf{ITW}} & \multicolumn{1}{c}{\textbf{MLAAD}} & \multicolumn{1}{c}{\textbf{TIMIT-TTS}} & \multicolumn{1}{c}{\textbf{WaveFake}} & \multicolumn{1}{c}{\textbf{OOD}} \\
\midrule
1                                  & \cellcolor[HTML]{F2A66D}5.8        & \cellcolor[HTML]{F3A86C}19.0       & \cellcolor[HTML]{F8BA6A}19.4              & \cellcolor[HTML]{F09F6E}38.0             & \cellcolor[HTML]{FFD466}15.8       & \cellcolor[HTML]{E67C73}86.7           & \cellcolor[HTML]{F3A86C}32.2          & \cellcolor[HTML]{EE976F}35.2     \\
2                                  & \cellcolor[HTML]{FAC169}3.0        & \cellcolor[HTML]{F3AB6C}18.2       & \cellcolor[HTML]{F8BB69}18.7              & \cellcolor[HTML]{EC926F}42.5             & \cellcolor[HTML]{F8BB69}21.4       & \cellcolor[HTML]{EB8D70}76.6           & \cellcolor[HTML]{EB8D70}43.1          & \cellcolor[HTML]{ED936F}36.8     \\
3                                  & \cellcolor[HTML]{FCC968}2.3        & \cellcolor[HTML]{F5B06B}16.7       & \cellcolor[HTML]{FBC668}13.6              & \cellcolor[HTML]{F1A26D}37.0             & \cellcolor[HTML]{FFD666}15.2       & \cellcolor[HTML]{F3AB6C}58.7           & \cellcolor[HTML]{F3A96C}32.0          & \cellcolor[HTML]{F3AB6C}28.9     \\
4                                  & \cellcolor[HTML]{FFD666}1.0        & \cellcolor[HTML]{F8BB69}13.1       & \cellcolor[HTML]{FED166}8.3               & \cellcolor[HTML]{FAC468}25.1             & \cellcolor[HTML]{FFD666}15.2       & \cellcolor[HTML]{F5B06B}55.8           & \cellcolor[HTML]{F9BF69}23.1          & \cellcolor[HTML]{F8BC69}23.4     \\
5                                  & \cellcolor[HTML]{CFCE71}0.7        & \cellcolor[HTML]{FCCB67}8.2        & \cellcolor[HTML]{E8D26B}5.2               & \cellcolor[HTML]{ABC878}12.8             & \cellcolor[HTML]{FFD466}15.7       & \cellcolor[HTML]{FAC169}45.4           & \cellcolor[HTML]{C9CD72}10.1          & \cellcolor[HTML]{FED266}16.2     \\
6                                  & \cellcolor[HTML]{87C280}0.4        & \cellcolor[HTML]{F2D369}4.5        & \cellcolor[HTML]{F7D468}5.4               & \cellcolor[HTML]{6FBE85}8.4              & \cellcolor[HTML]{FED266}16.2       & \cellcolor[HTML]{C7CD72}23.3           & \cellcolor[HTML]{70BF85}4.4           & \cellcolor[HTML]{93C47D}10.4     \\
7                                  & \cellcolor[HTML]{E7D26C}0.8        & \cellcolor[HTML]{BECB74}4.1        & \cellcolor[HTML]{C9CD72}4.8               & \cellcolor[HTML]{57BB8A}6.6              & \cellcolor[HTML]{E9D26B}14.4       & \cellcolor[HTML]{A2C77A}17.2           & \cellcolor[HTML]{60BC88}3.4           & \cellcolor[HTML]{63BD88}8.4      \\
8                                  & \cellcolor[HTML]{87C280}0.4        & \cellcolor[HTML]{A4C77A}3.9        & \cellcolor[HTML]{FFD666}5.6               & \cellcolor[HTML]{95C57D}11.2             & \cellcolor[HTML]{87C280}10.8       & \cellcolor[HTML]{FFD666}32.6           & \cellcolor[HTML]{84C281}5.7           & \cellcolor[HTML]{B3C977}11.6     \\
9                                  & \cellcolor[HTML]{9FC67B}0.5        & \cellcolor[HTML]{7DC182}3.6        & \cellcolor[HTML]{8CC37F}4.0               & \cellcolor[HTML]{B1C977}13.2             & \cellcolor[HTML]{7AC083}10.3       & \cellcolor[HTML]{CECE71}24.5           & \cellcolor[HTML]{79C083}5.0           & \cellcolor[HTML]{8DC37F}10.1     \\
10                                 & \cellcolor[HTML]{B7CA76}0.6        & \cellcolor[HTML]{E5D16C}4.4        & \cellcolor[HTML]{75BF84}3.7               & \cellcolor[HTML]{CFCE71}15.4             & \cellcolor[HTML]{57BB8A}9.0        & \cellcolor[HTML]{DDD06E}27.0           & \cellcolor[HTML]{8FC47E}6.4           & \cellcolor[HTML]{A3C77A}11.0     \\
11                                 & \cellcolor[HTML]{FFD666}0.9        & \cellcolor[HTML]{B1C977}4.0        & \cellcolor[HTML]{7DC182}3.8               & \cellcolor[HTML]{E6D26C}17.1             & \cellcolor[HTML]{5CBB89}9.2        & \cellcolor[HTML]{FCD567}32.1           & \cellcolor[HTML]{84C281}5.7           & \cellcolor[HTML]{BBCB75}12.0     \\
12                                 & \cellcolor[HTML]{B7CA76}0.6        & \cellcolor[HTML]{E5D16C}4.4        & \cellcolor[HTML]{A3C77A}4.3               & \cellcolor[HTML]{FED266}20.3             & \cellcolor[HTML]{7FC182}10.5       & \cellcolor[HTML]{FFD666}32.7           & \cellcolor[HTML]{95C57D}6.8           & \cellcolor[HTML]{D9CF6F}13.2     \\
13                                 & \cellcolor[HTML]{E7D26C}0.8        & \cellcolor[HTML]{FFD666}4.9        & \cellcolor[HTML]{E0D16D}5.1               & \cellcolor[HTML]{FED266}20.6             & \cellcolor[HTML]{98C57D}11.4       & \cellcolor[HTML]{FDCE67}37.7           & \cellcolor[HTML]{A8C879}8.0           & \cellcolor[HTML]{FCD567}14.6     \\
14                                 & \cellcolor[HTML]{FFD666}0.9        & \cellcolor[HTML]{FFD566}5.2        & \cellcolor[HTML]{FFD666}5.5               & \cellcolor[HTML]{FED066}21.1             & \cellcolor[HTML]{ABC878}12.1       & \cellcolor[HTML]{FED366}34.8           & \cellcolor[HTML]{C0CB74}9.5           & \cellcolor[HTML]{FFD666}14.7     \\
15                                 & \cellcolor[HTML]{FFD666}1.0        & \cellcolor[HTML]{FFD566}5.2        & \cellcolor[HTML]{FFD666}5.5               & \cellcolor[HTML]{FFD566}19.4             & \cellcolor[HTML]{D6CF6F}13.7       & \cellcolor[HTML]{FFD566}33.7           & \cellcolor[HTML]{FFD566}14.0          & \cellcolor[HTML]{FFD566}15.3     \\
16                                 & \cellcolor[HTML]{FED166}1.5        & \cellcolor[HTML]{FFD566}5.0        & \cellcolor[HTML]{FFD666}5.6               & \cellcolor[HTML]{EBD26B}17.5             & \cellcolor[HTML]{A0C67B}11.7       & \cellcolor[HTML]{D7CF6F}26.0           & \cellcolor[HTML]{B9CA75}9.1           & \cellcolor[HTML]{C8CD72}12.5     \\
17                                 & \cellcolor[HTML]{FDCD67}1.8        & \cellcolor[HTML]{FFD666}4.6        & \cellcolor[HTML]{FFD666}5.5               & \cellcolor[HTML]{FFD666}18.9             & \cellcolor[HTML]{D9CF6F}13.8       & \cellcolor[HTML]{BDCB75}21.7           & \cellcolor[HTML]{FBC768}19.8          & \cellcolor[HTML]{EED36A}14.1     \\
18                                 & \cellcolor[HTML]{FFD566}1.1        & \cellcolor[HTML]{F2D369}4.5        & \cellcolor[HTML]{EFD36A}5.3               & \cellcolor[HTML]{FFD466}19.7             & \cellcolor[HTML]{B8CA76}12.6       & \cellcolor[HTML]{AFC978}19.3           & \cellcolor[HTML]{F5D469}12.9          & \cellcolor[HTML]{C5CC73}12.4     \\
19                                 & \cellcolor[HTML]{FBC668}2.5        & \cellcolor[HTML]{A4C77A}3.9        & \cellcolor[HTML]{8CC37F}4.0               & \cellcolor[HTML]{97C57D}11.3             & \cellcolor[HTML]{DAD06E}13.9       & \cellcolor[HTML]{8DC37F}13.7           & \cellcolor[HTML]{FDCE67}16.8          & \cellcolor[HTML]{99C57C}10.6     \\
20                                 & \cellcolor[HTML]{FFD566}1.1        & \cellcolor[HTML]{57BB8A}3.3        & \cellcolor[HTML]{66BD87}3.5               & \cellcolor[HTML]{7FC182}9.6              & \cellcolor[HTML]{FFD366}16.0       & \cellcolor[HTML]{7BC083}10.6           & \cellcolor[HTML]{DCD06E}11.3          & \cellcolor[HTML]{73BF84}9.1      \\
21                                 & \cellcolor[HTML]{FFD466}1.1        & \cellcolor[HTML]{57BB8A}3.3        & \cellcolor[HTML]{D1CE70}4.9               & \cellcolor[HTML]{BACB75}13.9             & \cellcolor[HTML]{D9CF6F}13.8       & \cellcolor[HTML]{79C083}10.3           & \cellcolor[HTML]{92C47E}6.6           & \cellcolor[HTML]{6DBE86}8.8      \\
22                                 & \cellcolor[HTML]{FED266}1.3        & \cellcolor[HTML]{8AC37F}3.7        & \cellcolor[HTML]{D8CF6F}5.0               & \cellcolor[HTML]{B9CA75}13.8             & \cellcolor[HTML]{F4D469}14.8       & \cellcolor[HTML]{57BB8A}4.6            & \cellcolor[HTML]{FFD666}13.5          & \cellcolor[HTML]{77C083}9.2      \\
23                                 & \cellcolor[HTML]{FDCC67}1.9        & \cellcolor[HTML]{A4C77A}3.9        & \cellcolor[HTML]{C1CC74}4.7               & \cellcolor[HTML]{C9CD72}15.0             & \cellcolor[HTML]{F1D369}14.7       & \cellcolor[HTML]{8BC37F}13.3           & \cellcolor[HTML]{F9BE69}23.3          & \cellcolor[HTML]{C8CD72}12.5     \\
24                                 & \cellcolor[HTML]{FAC468}2.8        & \cellcolor[HTML]{FFD566}5.0        & \cellcolor[HTML]{FED266}7.7               & \cellcolor[HTML]{EBD26B}17.5             & \cellcolor[HTML]{F2A66C}26.2       & \cellcolor[HTML]{A6C779}17.9           & \cellcolor[HTML]{F6B56A}27.0          & \cellcolor[HTML]{FED066}16.9   \\
\bottomrule

\end{tabular}
     \caption{%
     Wav2Vec2-300m layer-wise results on testing datasets.
     Each column represents EER(\%) results on each dataset.
     The OOD column represents the mean EER over out-of-domain datasets.
     The color ranges are normalised per column.
    }
    \label{tab:layer_results}
\end{table}

\begin{figure}
    \centering
    \includegraphics[width=0.7\linewidth]{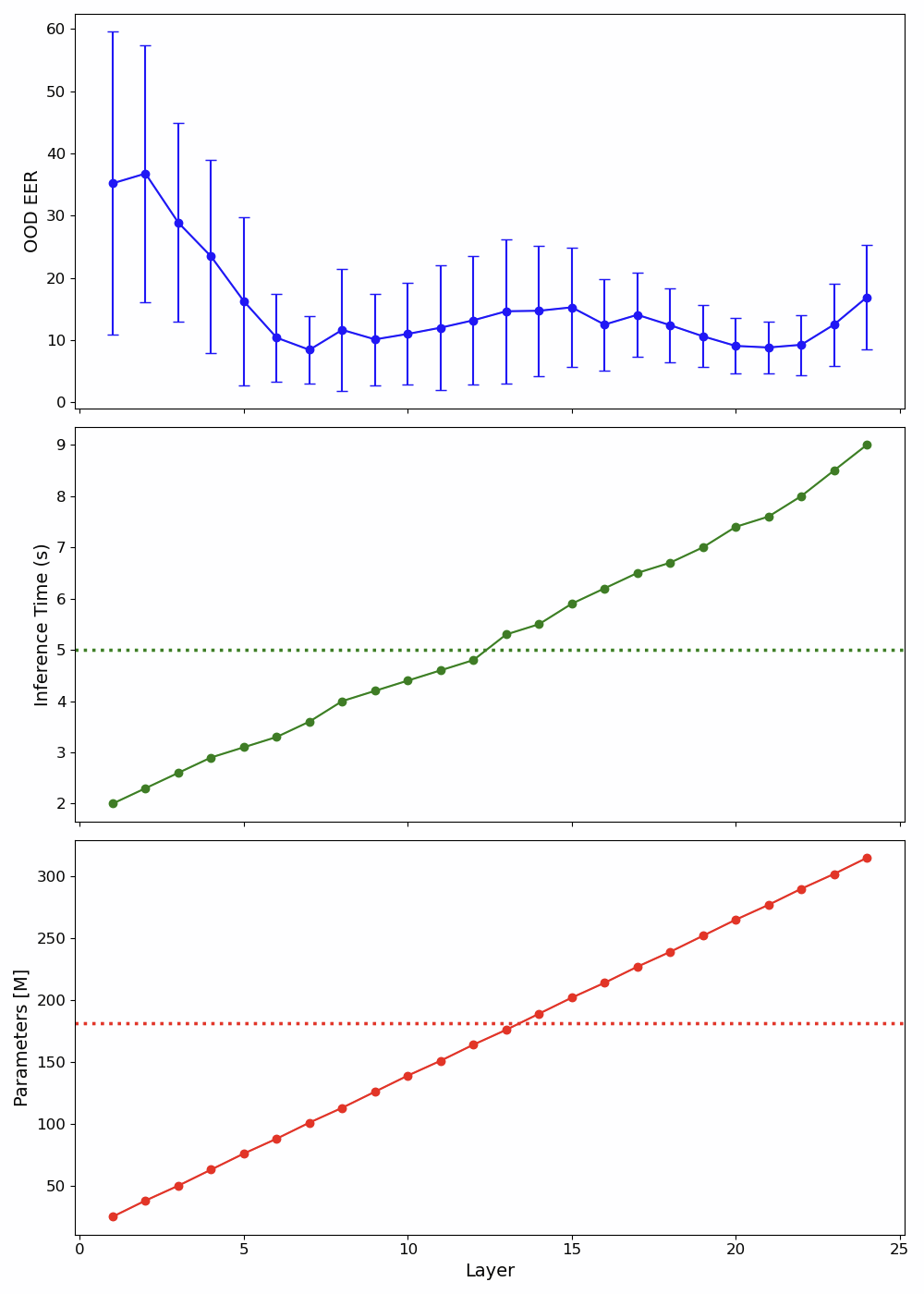}
    \caption{Tradeoff between Out-Of-Domain performance, inference time and memory footprint for the various Wav2Vec2-300m layers.} 
    \label{fig:layer_eer}
\end{figure}

To evaluate the generalization capabilities of the \textit{Wav2Vec2-300m} self-supervised learning (SSL) model, we conducted experiments testing each individual layer's performance across multiple out-of-domain datasets.
The model was trained exclusively on the ASVspoof19 (ASV19) training partition and evaluated on ASVspoof21 DF (ASV21), Fake or Real (FoR), In-the-Wild (ITW), MLAAD, TIMIT-TTS, and WaveFake. The Equal Error Rate (EER) was used as the primary evaluation metric.  

\mypar{Performance.} Table \ref{tab:layer_results} presents the layer-wise Equal Error Rate (EER, \%) across various testing datasets using the \textit{Wav2Vec2-300m} model.
Each column corresponds to a specific dataset, while each row indicates the EER at a specific transformer layer within the model.
The last column (OOD) reports the mean EER over the out-of-domain datasets, that is, all datasets except ASV19, which was used for training.

Overall, the results reveal a clear trend in model performance across layers.
The EER consistently decreases throughout the initial layers, reaching a broad performance optimum between layers 5 and 7.
After this point, the EER gradually increases again as we move toward deeper layers, peaking around layers 13-15 for several datasets. 
A secondary performance peak is observed around layers 19-21, suggesting that some deeper representations still retain discriminative features for certain scenarios, particularly under the out-of-domain evaluation.

However, the optimal layer---the one achieving the lowest EER---varies significantly across datasets:
(1) ASV19 achieves its lowest EER at layer 6 (0.4\%), indicating strong discriminative power in the mid-layers for speaker verification.
(2) ASV21 follows a similar trend, with a minimum at layer 9 (3.6\%). 
(3) FoR reaches the optimal performance at layer 10 (3.7\%), with slightly worse performance in both shallower and deeper layers.
(4) ITW exhibits a more gradual decline, with its best performance at layer 7 (6.6\%), but less dramatic variation across layers.
(5) MLAAD shows a noisier trend, but reaches its minimum at layer 10 (9.0\%), suggesting that mid-to-late layers capture useful artifacts for this dataset.
(6) TIMIT-TTS improves steadily until layer 22 (4.6\%), then slightly worsens.
(7) WaveFake reaches a sharp minimum at layer 7 (3.4\%), then degrades significantly in deeper layers, highlighting the importance of mid-level representations for this dataset.

The mean OOD performance is lowest at layer 7 (8.4\%), confirming that early intermediate layers yield the strong representations for cross-dataset generalization.

\mypar{Performance and efficiency as a function of layer.}
By truncating the feature extractor at different levels, we influence not only the predictive performance, but crucially also the computational efficiency.
From a computational perspective, we measure the inference time and the memory footprint.
Figure \ref{fig:layer_eer} shows three plots illustrating these three metrics: the detection performance, the inference time, and the number of parameters per layer.
 
The top plot depicts visually the results in Table~\ref{tab:layer_results}.
The plot shows the mean OOD performance together with the standard deviation
(the mean and standard deviation are computed across the six OOD datasets).
We can neatly see the two optima in performance, around layers 7 and 21, respectively.

The middle plot tracks inference time across layers.
Inference is done on a 5s audio sampled at 16 kHz and using a single CPU core.
The horizontal dotted line indicates a real time factor of one, that is, predicting on an audio takes as much time as that audio's duration.
The plot shows a clear linear increase in inference time as the number of transformer layers grows, starting at roughly 2 seconds for early layers and reaching approximately 9 seconds for the full model.
This scaling reinforces the computational cost of deeper layers.
In addition, this result indicates that the full \textit{Wav2Vec2} model cannot be used for real-time processing on the single CPU core available to the browser extension. 

The bottom plot indicates the number of parameter count when truncating the model at different depths.
Similar to the inference time, the parameter count grows linearly, starting around 30 million at early layers and exceeding 300 million by the final layer.
In order to deploy a Chrome extension on the Chrome Web Store, there is a hard requirement of using less than 2 GB of storage.
The dotted line represents the cutoff point at which the model uses 2 GB of storage, corresponding to 181M parameters.

The main takeaway is that layer 7 offers the best detection performance, while respecting the timing and memory requirements for a browser extension.

\subsection{Comparison to other approaches}
Going further we compared our approach with state-of-the-art audio deepfake detection models, including \textit{RawGAT \cite{tak2021end}, RawNet2 \cite{tak2020end}, RawNet3\cite{jung2022pushing} }, and \textit{AASIST} \cite{tak2022automatic}. All the models were trained on the ASV19  training partition and evaluated on multiple out-of-domain (OOD) datasets, including ASV21, ITW, TIMIT-TTS, FoR, MLAAD, and WaveFake. The performance of each model was measured using Equal Error Rate (EER), where lower values indicate better generalization. The results are summarized in Table \ref{tab:archi-comparison}.

\begin{table*}[h]
    \centering
    \begin{tabular}{l|rrrrrrr|r}
        \toprule
        Model  & \multicolumn{1}{c}{ASV19} & \multicolumn{1}{c}{ASV21} & \multicolumn{1}{c}{ITW} & \multicolumn{1}{c}{TIMIT-TTS} & \multicolumn{1}{c}{FoR} & \multicolumn{1}{c}{MLAAD} & \multicolumn{1}{c|}{WaveFake} & \multicolumn{1}{c}{OOD} \\
        \midrule
        RawNet2 \cite{tak2020end}  & 4.1 & 22.2 & 33.7 & \textbf{7.8} & 31.0 & 12.4 & 36.5 & 23.9 \\
        RawNet3 \cite{jung2022pushing}  & 6.2 & 24.8 & 29.7 & 22.3 & 63.3 & 56.4 & 40.8 & 39.6 \\
        W2V2-layer24  & 2.8 & 5.0 & 7.7 & 17.5 & 26.2 & 17.9 & 27.0 & 16.9 \\
        RawGAT \cite{tak2021end} & 1.2 & 18.5 & 33.8 & 25.8 & 33.4 & 18.1 & 32.1 & 27.0 \\
        AASIST \cite{tak2022automatic} & \textbf{0.2} & 7.3 & 11.2 & 15.6 & \textbf{3.9} & \textbf{10.9} & 18.8 & 11.3 \\
        W2V2-layer7 (ours)  &  0.8 & \textbf{4.1} & \textbf{6.6} & 17.2 & 4.8 & 14.4 & \textbf{3.4} & \textbf{8.4} \\
        \bottomrule
    \end{tabular}
    \caption{Performance comparison of different models. The bolded values indicate the lowest error per column.
    }
    \label{tab:archi-comparison}
\end{table*}

\mypar{Out-of-domain performance.}
The \textit{Wav2Vec2-300m}  truncated model at layer 7 (W2V2-layer7) achieved the lowest average EER (8.4\%) across out-of-domain datasets, demonstrating superior generalization compared to other deepfake detection models. Notably, it outperformed \textit{AASIST} with fine-tuned \textit{Wav2Vec2-300m} (11.3\%) and \textit{RawGAT} (27.0\%), highlighting the effectiveness of mid-level SSL representations for deepfake detection.

Among non-SSL-based models, \textit{RawNet2} and \textit{RawNet3} struggled with generalization, yielding average EERs of 23.9\% and 39.6\%, respectively.
While \textit{RawNet2} performed well on ASV19, its generalization to more diverse datasets (e.g., ASV21, ITW, FoR, WaveFake) was significantly weaker.
Similarly, \textit{RawGAT} exhibited competitive performance on ASV19 (1.2\%) but failed to generalize well, reaching 33.8\% EER on ITW and 32.1\% on WaveFake.

\begin{figure}
    \centering
    \includegraphics[width=0.7\linewidth]{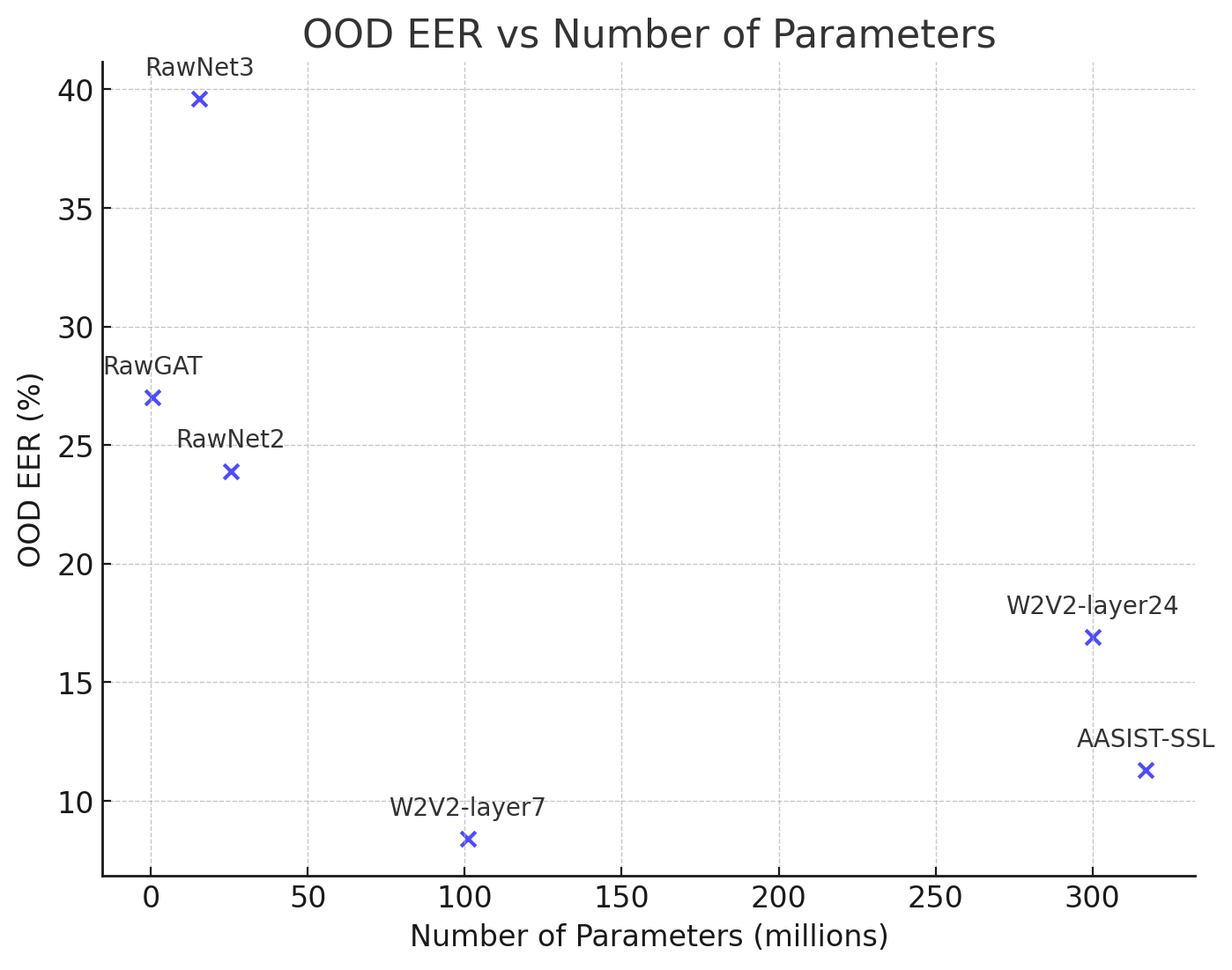}
    \caption{Out-of-domain EER vs model parameters for tested architectures}
    \label{fig:perf-vs-dim}
\end{figure}

\mypar{Computational efficiency.} 
Apart from out-of-domain performance,
computational efficiency plays a critical role in determining the practical usability of deepfake detection models. This is particularly relevant for real-time applications, privacy-preserving deployment, and resource-constrained environments such as browser extensions or mobile applications. 
Figure \ref{fig:perf-vs-dim} summarizes the parameter count of each model alongside their out-ow-domain Equal Error Rates (EERs). 

Among the evaluated models, \textit{RawGAT} (400k parameters) is the smallest,wmaking it lightweight enough for such applications, but its high OOD EER (27\%) renders it unreliable. Conversely, the truncated \textit{Wav2Vec2-300m} (layer 7) model (101M parameters, 8.4\% EER) presents a viable compromise, maintaining high accuracy while remaining computationally efficient enough for browser-based inference. We can observe that \textit{RawNet3} and \textit{Wav2Vec2-300m} untruncated (W2V2-layer24) do not offer advantages in terms of OOD EER or number of parameters compared to the other tested models.

\mypar{Different use cases for different model sizes.}
Depending on the application, different models would be preferable from the ones tested.
In the case of extreme resource-constrained environments (e.g. IoT devices, embedded systems),  \textit{RawGAT} (400k, 27\% EER) would be the most suited but performs poorly in out-of-domain settings.
For cloud-based, high-performance detections (e.g. forensic analysis, large-scale platform moderation) inference latency and computational resources might not be as important as overall performance, in which case \textit{AASIST} (317M, 11.3\% EER) would be a viable solution.
As for real-time, on-device applications (e.g. Chrome extensions, mobile apps) where there are also resource constraints but not as extreme as in the first case, \textit{Wav2Vec2-300m} truncated at layer 7 (101M, 8.4\% EER) offers the best trade-off between accuracy and efficiency.
\textit{RawNet2} (25.4M, 23.9\% EER) is also an option, being more lightweight, but lacking generalization power.

\section{Chrome Extension for Deepfake Detection}

To make deepfake detection accessible, we developed a Chrome extension that runs the truncated \textit{Wav2Vec2-300m} model for audio verification.
The extension predicts if an audio is fully spoofed or not, and returns the final decision.

\mypar{Pipeline.} In order to produce the results, the extension has the following properties (Figure \ref{fig:UI}):
(1) The first 5 seconds of the audio are cut and used for the prediction.
(2) The truncated \textit{Wav2Vec2-300m} model at the best performing layer is used, taking into account our computational analysis in Section~\ref{subsec:layerwise-analysis}.
(3) We use the ONNX Runtime to ensure lightweight model execution.
(4) The model runs entirely in-browser, requiring only a single CPU core.
(5) The extension outputs the prediction: bonafide or spoofed.

\mypar{Deployment and accessibility.} We packaged the extension for Chrome deployment. Local audio files can be used directly with the extension, ensuring privacy as no information is transmitted through a server. In order to run the extension, the package must be added to Chrome using the \texttt{chrome://extensions} tab with the Developer Tools option activated.
In order to run inference, the package uses 1GB of RAM.

\mypar{Performance testing.} We tested detection on samples from the testing datasets and ensured that the results are identical to those from the Python environment.

In order to test performance, we used the following hardware environments:
(1) Ryzen 7 7800$\times$3d CPU: This was also used as the developing environment, processing a 5 seconds audio using the Chrome extension takes 3.4 seconds on a single core.
(2) In order to simulate a common laptop CPU we also tested on Intel i3-1215U.
This time, processing a 5 seconds audio using the chrome extension took 4.2 seconds on a single CPU core.

\section{Conclusions}

In this study, we systematically evaluate self-supervised learning (SSL) models for audio deepfake detection, comparing them against state-of-the-art deepfake detection architectures such as \textit{RawNet2, RawNet3, RawGAT}, and \textit{AASIST}. Our analysis focuses on three key aspects: generalization, computational efficiency, and real-world deployability.  

We benchmark across 6 different out of domain datasets which contain diverse Text-To-Speech and Vocoder architectures. Our results demonstrate that truncating \textit{Wav2Vec2-300m} at layer 7 significantly improves generalization and efficiency, achieving an out-of-domain Equal Error Rate (EER) of 8.4\%, outperforming larger models such as full \textit{Wav2Vec2-300m} (16.9\%), \textit{AASIST} (11.3\%), and traditional CNN-based architectures like \textit{RawNet3} (39.6\%). 
These findings suggest that mid-layer representations in SSL models retain highly discriminative features for detecting synthetic speech while reducing computational overhead. 

Furthermore, our study highlights the computational advantages of model truncation. The truncated \textit{Wav2Vec2-300m} model (101M parameters) provides a significantly lower memory footprint and faster inference speed than the full model (300M parameters), making it more suitable for real-time applications such as on-device detection, browser extensions, and mobile security applications. Compared to \textit{RawGAT} (400k parameters) and \textit{RawNet}-based models, the truncated SSL model balances high accuracy with practical feasibility, making it an ideal choice for privacy-preserving and energy-efficient deepfake detection.  

Finally, the proposed model was integrated into a browser extension that performs inference directly on local audio files, eliminating the need to upload data to external servers. This approach offers a practical and privacy-preserving solution for on-device deepfake detection.

\bibliographystyle{ieeetr}
\bibliography{mybib}
\end{document}